\documentclass[aps,prd,twocolumn,nofootinbib,superscriptaddress,amsmath,amssymb,floatfix,10pt]{revtex4-1}

\usepackage[T1]{fontenc}
\DeclareUnicodeCharacter{2297}{\otimes}
\usepackage{graphicx} 
\usepackage{dcolumn} 
\usepackage{tabularx}
\usepackage{amssymb} 
\usepackage{xcolor}
\usepackage{bm}      
\usepackage[compact]{titlesec}
\usepackage{booktabs}

\usepackage{natbib}
\usepackage{adjustbox}
\usepackage{multirow}
\usepackage[title]{appendix}

\def\emri#1{Extreme Mass-Ratio Inspiral#1 (EMRI#1)\gdef\emri{EMRI}}
\def\imbh#1{Intermediate Mass Black Hole#1 (IMBH#1)\gdef\imbh{IMBH}}
\def\smbh#1{supermassive black hole#1(SMBH#1)\gdef\smbh{SMBH}}
\def\bbh#1{binary black hole#1 (BBH#1)\gdef\bbh{BBH}}
\def\pbh#1{primordial black hole#1 (PBH#1)\gdef\pbh{PBH}}
\def\imbhb#1{intermediate mass black hole binary#1 (IMBHB#1)\gdef\imbhb{IMBHB}}
\def\hmns#1{hypermassive neutron star#1 (HMNS#1)\gdef\hmns{HMNS}}
\def\bh#1{black hole#1 (BH#1)\gdef\bh{BH}}
\def\ns#1{neutron star#1 (NS#1)\gdef\ns{NS}}
\def\hmns#1{hyper-massive neutron star#1 (HMNS#1)\gdef\hmns{HMNS}}
\def\nsbh#1{neutron star-black hole#1 (NSBH#1)\gdef\nsbh{NSBH}}
\def\bns#1{binary neutron star#1 (BNS#1)\gdef\bns{BNS}}
\def\gw#1{gravitational wave#1 (GW#1)\gdef\gw{GW}}
\def\eos#1{equation of state#1 (EOS#1)\gdef\eos{EOS}}
\def\gpu#1{graphics processing unit#1 (GPU#1)\gdef\gpu{GPU}}
\def\gr#1{General Relativity#1 (GR#1)\gdef\gr{GR}}
\def\cbc#1{Compact Binary Coalescence#1 (CBC#1)\gdef\cbc{CBC}}
\def\eob#1{Effective-One-Body#1 (EOB#1)\gdef\eob{EOB}}
\def\pnw#1{post-Newtonian#1 (PN#1)\gdef\pnw{PN}}
\def\pmw#1{post-Minkowskian#1 (PM#1)\gdef\pmw{PM}}
\def\hom#1{Higher Order Mode#1 (HOM#1)\gdef\hom{HOM}}
\def\agn#1{Active Galactic Nuclei#1 (AGN#1)\gdef\agn{AGN}}

\def\ligo#1{Laser Interferometer Gravitational-Wave Observatory#1 (LIGO#1)\gdef\ligo{LIGO}}
\def\lvk#1{LIGO-Virgo-KAGRA#1 (LVK#1)\gdef\lvk{LVK}}

\def\lisa#1{Laser Interferometer Space Antennae#1 (LISA#1)\gdef\lisa{LISA}}

\newcommand{\pycbc}{{PyCBC}}

\newcommand{\bilby}{{Bilby}}
\newcommand{\gwpy}{{GWpy}}

\newcommand{\lal}{{\texttt{LALSuite}}}

\usepackage{float}

\begin{document}

\title{Algorithm for Dark Matter-Admixed Neutron Stars}

\author{Nguyen Thi Lan Anh, Peter Lott, and Quynh Lan Nguyen}
\email{lan.nguyenquynh@phenikaa-uni.edu.vn}
\affiliation{Phenikaa Institute for Advanced Study, Phenikaa University, Duong Noi, 12116 Hanoi, Vietnam}

\date{\today}
\begin{abstract}
Gravitational-wave observations provide a unique window into the fundamental nature of massive objects. In particular, neutron star equations of state have been constrained due to the success of gravitational wave observatories. Recently, the possibility of detecting dark matter-admixed neutron stars via ground-based laser interferometry has been explored. Dark matter would impact the gravitational waveform of an inspiraling neutron star system through tidal parameters, namely the tidal deformability $\lambda$, incurring a phase shift to the frequency evolution of the signal. This phase shift would depend both on the percentage of dark matter within the star and its particle nature, e.g., bosonic or fermionic. Indirect detection of dark matter through admixture within neutron stars can provide insight into the neutron equation of state, as well as constraints on the density of dark matter in the universe. In this work, we introduce \texttt{Darksuite}, a proposed extension of the \lal{} software framework, designed to model the gravitational wave signatures of dark-matter-admixed neutron stars. This framework employs simulations from the two-fluid, generally relativistic Tolman-Oppenheimer-Volkoff equations, wherein one fluid is ordinary nuclear matter and the other is dark matter. We demonstrate interpolation of values from a bank of simulations, enabling the study of binary systems where at least one component may be a dark-matter-admixed neutron star.  By leveraging existing methodologies within \lal{} for tidal phase corrections and supplementing them with dark matter effects, \texttt{Darksuite} provides a means to generate and analyze gravitational waveforms for these exotic systems.

\end{abstract}
\maketitle
\section{Introduction}
The existence of dark matter is a cornerstone of modern cosmology, supported by a wide range of astrophysical observations. Evidence from galactic rotation curves~\cite{Rubin1970}, gravitational lensing~\cite{Clowe2006}, cosmic microwave background anisotropies~\cite{Planck2018}, and large-scale structure formation~\cite{White1978} all point to a nonluminous form of matter and energy that constitutes approximately 95\% of the total matter and energy content in the universe ~\cite{doi:10.1142/S0217751X13300421}. Despite its significant gravitational influence, dark matter has eluded direct detection, and its fundamental nature remains one of the most pressing open questions in physics.

Binary neutron star (BNS) systems with an admixture of dark matter within could provide one with an avenue for indirect detection of dark matter ~\cite{Leung:2022wcf, Nelson_2019}.  Neutron stars (NSs) are incredibly dense, and their nuclear core may provide an ideal laboratory for the bombardment of dark matter. \lvk{} collaboration has detected several \gw{} signals from binary mergers, several of them associated with the BNS or \nsbh{} systems. The first confident BNS event, GW170817 ~\cite{GW170817}, provided combined detection of the \gw{} signal with its electromagnetic counterpart~\cite{GW170817, Abbott2017_MMA}. The first confident NSBH detections, GW200105 and GW200115, involved black holes of $\sim 6\text{--}9 M_\odot$ and neutron stars of $\sim 1.5\text{--}2 M_\odot$, although no associated electromagnetic counterparts were observed, suggesting prompt neutron star engulfment \cite{Abbott2021_NSBH}. Additionally, the event GW190814 featured a secondary object with mass $\sim 2.6 M_\odot$, within the so-called mass gap, and may represent either a heavy neutron star or a light black hole \cite{Abbott2020_GW190814}. Thus, LVK has confidently reported detections of two BNS mergers, two NSBH mergers, and one ambiguous case potentially involving a neutron star. These observations provide constraints on the neutron star equation of state (EOS) --- which describes how matter behaves at the extreme densities inside neutron stars --- merger rates, and the compact object mass function, and mark the beginning of systematic population studies of the BNS and NSBH systems \cite{Abbott2021_O3bCat}.

The constraints on neutron star tidal parameters---quantities that describe how neutron stars deform under the influence of their companion’s gravitational field--are strongly dependent on the (signal-to-noise ratio) SNR at high frequencies (typically $> 400$ Hz), where the tidal effects influence phase evolution during the late inspiral. Specifically, tidal deformability, which is a measure of how easily a neutron star shape is distorted by external gravitational fields, can be restricted in BNS mergers, which typically have long inspirals and strong tidal interactions from both stars. In contrast, NSBH mergers generally show minimal tidal signatures because there is only one neutron star in the system. Moreover, when the neutron star is promptly engulfed by the black hole, as suggested by the absence of electromagnetic counterparts in events like GW200105 and GW200115, the phase shift due to tidal effects becomes even smaller, reducing the ability to constrain ${\lambda}$.

Since parameter estimation of NS has been done using \gw{} templates, it may be feasible to constrain dark matter content within a candidate neutron star. \texttt{Darksuite} is a proposed extension of \lal{} designed to incorporate light dark matter effects into existing gravitational waveform models. This extension aims to facilitate the analysis of gravitational wave signals from binary systems involving neutron stars, such as the NS-NS (neutron star-neutron star) and NSBH (neutron star-black hole) systems, in which at least one component is a dark matter-admixed neutron star. The waveform of such systems is mainly influenced by the tidal Love number, $k_2$, which characterizes the response of a neutron star to an external tidal field. Another important parameter is the ratio of dark matter to ordinary baryonic matter, $f$, which, along with the assumed dark matter properties, dictates the impact of dark matter on the structure of the neutron star and its corresponding gravitational waveform.  

Theoretical models, such as those in \cite{Leung:2022wcf}, provide a framework for computing $k_2$ under different dark-matter prescriptions, considering scenarios where dark matter behaves either as an ideal bosonic gas or as an ideal fermionic gas. In parallel, the tidal phase correction methodology outlined by \cite{Dietrich:2017aum} has been implemented in \lal{} through models such as \texttt{NRTidal}, \texttt{NRTidalv2}, and \texttt{NRTidalv3}, offering a foundation for extending waveform models to include dark matter effects. In general, a dark matter particle obeying Bose-Einstein or Fermi-Dirac statistics could also be self-interacting \cite{lan06, doi:10.1142/S0217732321300019}. 

The rest of this article is organized as follows. In Section II, we describe the theoretical framework upon which the \texttt{ Darksuite } algorithm is based. In Section III, we outline the methodology, including how \texttt{ Darksuite} is integrated into the \lal{} infrastructure. Section IV highlights some of the results of our work. Finally, in section V, we conclude with a discussion on the applications of this work and the detectability of dark-matter-admixed NS systems. We use natural units consistently throughout this paper, which sets $c = G = 1$.

\section{Theoretical Background}
\subsection{Waveform}
In this section, we provide the theoretical framework under which we produce the waveforms, following the formalism laid out in~\cite{Dietrich:2017aum}. We consider the gravitational wave strain, $h$, which is generally a function of the thirteen parameters: binary masses $m_1$ and $m_2$, the spins of each star $\vec{a}_1$ and $\vec{a}_2$ (three components for each star), the sky location parameters right ascension $\alpha$, declination $\delta$ and luminosity distance $d_L$, inclination $\iota$, polarization angle $\psi$, and phase angle $\phi$.

In the case of neutron stars, more information must be computed to account for their structure. A spherical neutron star in the presence of a tidal field $\mathcal{E}_{ij}$ will have induced on it a quadrupole moment $Q_{ij}$; i.e., the neutron star will become oblong due to the external field. The tidal deformability $\lambda$ is the ratio of the external tidal field $\mathcal{E}_{ij}$ to the induced quadrupole moment of the neutron star $Q_{ij}$:

\begin{equation}
    \lambda = \frac{Q_{ij}}{\mathcal{E}_{ij}},
\end{equation}

so in the case of \bns{s}, the two tidal deformabilities due to the stars' mutual attraction $\lambda_1$ and $\lambda_2$, must be included, for a total of fifteen parameters.
The effect of $\lambda_1$ and $\lambda_2$ on the waveform is to add a phase change $\delta \Psi$. In the Fourier domain, $h(f)$, the Fourier transform of the waveform may be decomposed into its amplitude and phase components~\cite{Dietrich:2017aum}.
\begin{equation}
    h(f) = A (f) e^{i \phi},
\end{equation}

where $A(f)$ is the amplitude of the gravitational wave, and $\phi = \Psi + \delta \Psi$ is the phase, with $\Psi$ the unperturbed phase and $\delta \Psi$ the phase shift due to tidal deformation. For simplicity, we adopt the functional form for the phase shift computed by~\cite{Dietrich:2017aum}, which is also implemented in the \texttt{NRTidal} family of waveforms. In the frequency domain, the phase shift is given by:
\begin{equation}
    \delta \Psi = - \frac{9}{16} \frac{v^5}{\mu M^4} \left[ \left( 11 \frac{m_2}{m_1} + \frac{M}{m_1}\right)\lambda_1 +  \left( 11 \frac{m_1}{m_2} + \frac{M}{m_2}\right)\lambda_2\right],
\end{equation}
where $v = \left(\pi M f\right)^{1/3}$ is a parameter that involves the frequency $f$ and the total mass $M$. 
The effect of tidal parameters on \bns{} systems can be visualized using tools like \lal{}, a publicly available package for simulating \gw{s} from compact objects.

\subsection{TOV equations in two-fluid framework}
We now turn to the description of dark-matter-admixed neutron stars. In these systems, the tidal deformability of each neutron star is influenced by the presence of dark matter in the companion star's core.
To capture the effects of the dark matter content within the stars, we solve the Tolman-Oppenheimer-Volkoff equations (TOV)\cite{PhysRev.55.374}, which govern the structure of spherically symmetric bodies undergoing hydrostatic equilibrium within a generally relativistic framework. Considering that the star consists of two fluids that only interact through gravity, the TOV equations (for the two types of fluids: dark matter (DM) and nuclear matter (NM)) must be modified accordingly:
\begin{align}
&\frac{dm_i}{dr} =  \;4\pi r^2 \rho_i, \\
&\frac{dp_i}{dr} = -\frac{m+4\pi r^3 p}{r^2(1-2m/r)}(\rho_i+p_i),\label{eq:dpdr}\\
&\frac{d\nu}{dr} = \frac{2 (m+4\pi r^3 p)}{r^2 (1 - 2m/r)} ,
\label{eq:TOV}
\end{align}
where $i$ denotes DM or NM, $\rho \equiv \rho_\text{NM}+ \rho_\text{DM}$ and $p \equiv p_\text{NM} + p_\text{DM}$ represent the energy density and pressure, respectively. So far, we only have three equations to solve for four unknowns; hence, the TOV equations must be supplemented with an additional equation - the equation of state (EOS) - which describes the relationship between pressure and density for a given type of matter. 
The boundary conditions are set at the center of the star, where \(m_{i,{(r=0)}} = 0\) and \(\rho_{i,{(r=0)}} = \rho_{i,c}\), with $\rho_{i,c}$ being the central energy density. The integration proceeds outward from \(r = 0\) to the stellar radius \(R\), defined by the condition \(p_\text{NM}(R) = p_\text{DM}(R) = 0\). 
The total mass of the star is given by \(M = m_\text{NM}(R_\text{NM}) + m_\text{DM}(R_\text{DM})\). 
After solving the TOV equations for a two-fluid star, we obtain two distinct radii: one of the dark matter components and one for the nuclear matter. The equatorial radius of the star is defined as the larger of the two~\cite{PhysRevD.109.023008}:
\[R = \max\{ R_{\text{NM}}, R_{\text{DM}} \}.\] 
This leads to configurations where the star may possess either a dark matter core or a dark matter halo, depending on the spatial distribution of the two components.

\subsection{Tidal deformability}
For practical purposes, we adopt the dimensionless form for the tidal deformability, $\Lambda \equiv \lambda / M^5$, or
\begin{equation}
\Lambda=\frac{2}{3}\frac{k_2}{C^5},  
\end{equation}
with $C=M/R$ being the dimensionless compactness parameter. The tidal Love number $k_2$ is given by~\cite{Hinderer_2008}: 
\begin{equation}
\begin{aligned}
    k_2 = &\, \frac{8}{5} C^5 \,( 1-2C)^2 \,\left[2-y_R + 2C(y_R-1) \right]\\ 
    & \times \{ 2C \, (6-3y_R + 3C\,(5y_R-8)) \\
    & + 4C^3 \left[13-11y_R + C\, (3y_R-2) + 2C^2 \, (1+y_R)\right]\\
    & +3\,(1-2C)^2 \left[2-y_R+2C\,(y_R-1)\right]\,\log{(1-2C)}\}^{-1} \\ 
\end{aligned}
\end{equation}
where $y_R$ is the metric function $y(r)$ evaluated at the stellar radius $R$, which is governed by the following equation:
\begin{equation}
    ry' + y^2 + y e^{\lambda(r)} \left[1 + 4 \pi r^2 \left( p - \rho \right)\right] + r^2 Q(r) = 0.
\end{equation}
where the prime denotes the derivative with respect to the radial variable $r$, and the quadrupole function $Q(r)$ is modified for the two-fluid formalism:
\begin{equation}
    Q(r) = 4 \pi e^{\lambda(r)} \left(5 \rho + 9 p + \sum_i \frac{\rho_i + p_i}{dp_i/d\rho_i}\right) - \frac{6 e^{\lambda(r)}}{r^2} - \nu'(r)^2
    \label{eqn:quad}
\end{equation}
with the metric function $\lambda(r)$ defined as:
\begin{equation}
    e^{\lambda(r)} = \left[1 - \frac{2m(r)}{r}\right]^{-1}
\end{equation}
The boundary condition for the metric variable $\nu(r)$ on the stellar surface is $e^{\nu} = 1 - 2C$, and for $y(r)$ in the center: $y(0) = 2$.

\subsection{Equations of State}
\subsubsection{Nuclear Matter}
In this work, we employ the Brussels-Montreal energy density functional BSk22 developed by Pearson et.al. \cite{bsk22} as a candidate to simulate the properties of nuclear matter under extreme conditions. This EOS was chosen because it provides a unified microscopic treatment of nuclear physics in all three regions of the neutron star (i.e. the outer crust, the inner crust and the core) using a single functional form. The model is based on a generalized Skyrme-type interaction, an effective approach widely used in nuclear astrophysics, and was precision-fitted to nearly all known atomic mass data. BSk22 can remain realistic at high densities and is capable of producing heavy neutron stars (up to 2.3 $M_\odot$),  in agreement with astrophysical observations. The EOS expression is reproduced as follows:

\begin{equation}
    \begin{aligned}
    \log_{10}(p)  &= \frac{a_1 + a_2\xi + a_3 \xi^3}{1+a_4 \xi} \{ \exp[a_5 (\xi-a_6) + 1\}^{-1}\\
    & +(a_7+ a_8 \xi)\{ \exp[a_9 (a_6-\xi)] + 1\}^{-1} \\
    & +(a_{10}+ a_{11} \xi)\{ \exp[a_{12} (a_{13}-\xi)] + 1\}^{-1} \\
    & +(a_{14}+ a_{15} \xi)\{ \exp[a_{16} (a_{17}-\xi)] + 1\}^{-1} \\
    & + \frac{a_{18}}{1+ [ a_{20} (\xi - a_{19})]^2} + \frac{a_{21}}{1+ [ a_{23} (\xi - a_{22})]^2}  \\
    \end{aligned}
    \label{eq:bsk22}
\end{equation} 
where $\xi \equiv \log_{10} (\rho / \text{g} \, \text{cm}^{-3})$. The functional includes 23 fitted parameters, listed in Table \ref{tab:bsk22}

\vspace{0.5 cm}

\begin{table}[ht]
    \centering
    \begin{tabular}{>{\centering\arraybackslash}m{0.1\linewidth} 
    >{\centering\arraybackslash}m{0.25\linewidth} 
    @{\hspace{0.03\linewidth}}||@{\hspace{0.03\linewidth}}
    >{\centering\arraybackslash}m{0.1\linewidth} 
    >{\centering\arraybackslash}m{0.25\linewidth}}
    $i$ & $a_i$ & $i$ & $a_i$ \rule[-2ex]{0pt}{3ex}\\
    \hline
    \rule{0pt}{3.5ex}
    1 & 6.682 & 13 & 14.135\\
    2 & 5.651 & 14 & 28.03\\
    3 & 0.00459 & 15 & -1.921\\
    4 & 0.14359 & 16 & 1.08\\
    5 &  2.681 & 17 & 14.89\\
    6 & 11.972  & 18 & 0.098\\
    7 & 13.993 & 19 & 11.67\\
    8 & 1.2904 & 20 & 4.75 \\
    9 &  2.665 & 21 & -0.037\\
    10 & -27.787 & 22 & 14.1\\
    11 & 2.014 & 23 & 11.9 \\
    12 & 4.09 & & 
    \end{tabular}
    \caption{Parameters $a_i$ of the Brussel-Montreal BSk22 functional from Equation \ref{eq:bsk22}.}
    \label{tab:bsk22}
\end{table} 
\subsubsection{Dark matter: Bosonic case}
We modeled the dark-matter component in an admixed neutron star using an EOS derived from a relativistic bosonic field with self-interaction. This model is from Flores et al. \cite{Flores_2019} covers a realistic parameter space for the self-coupling strength and the mass of the scalar boson, consistent with empirical constraints:
\[ 0.1 \,\mathrm{cm}^{2}/{\mathrm{g}}\leq\frac{\sigma}{m}\leq10 \,\mathrm{cm}^{2}/ \mathrm{g}.\]
The range has been widely recognized as astrophysically conceivable as it resolves small-scale structure issues in galaxies \cite{TULIN20181, Kaplinghat,Randall_2008}. For these reasons, we employ the following EOS to describe the bosonic self-interacting dark matter component: 
\begin{align}
p = \frac{4}{9}\rho_B[(1+\frac{3}{4}\rho/\rho_B)^{1/2}-1]^2 ,
\label{eq:boson_EOS}
\end{align}
where
\begin{align} 
\rho_B = \frac{\mu_B^4}{4a\hbar^3},
\end{align}
$\mu_B$ is the mass of bosonic dark matter, and
\(a\) is a dimensionless constant that describes the strength of the self-interaction.

In the low-density limit, the EOS reduces to the polytropic form:
\begin{align}
&p_\text{low} = \frac{1} {16\rho_B} \rho^2
\end{align}
while in the high-density limit, this EOS becomes ultra-relativistic:
\begin{align}
&p_{\text{high}} = \frac{1}{3}\rho.
\end{align}
\vspace{0.5 cm}
\subsubsection{Dark matter: Fermionic case}
The second participant for dark matter issued in this paper is the ideal Fermionic gas, which follows the parametric form for EOS \cite{fermi_eos}:
\begin{align}
\rho =& \:\rho_F(\sinh t - t),\\
p =& \:\frac{1}{3}\rho_F(\sinh t+3t-8\:\sinh\frac{t}{2}),
\end{align}
with
\begin{align}
    \rho_F &= \frac{1}{32\pi^2} \frac{\mu_F^4}{\hbar^3},\\
t &= 4\ln\left[\sqrt{1+\left(\frac{\hat{p}}{\mu_F}\right)^2} + \frac{\hat{p}}{\mu_F}\right],
\end{align}
where $\mu_F$ is the rest mass of the particle that obeys Fermi statistics and $\hat{p}$ is the maximum momentum related to the number density $n$ by:
\begin{align}
\hat{p} = \left(3 \pi^2 \hbar^3 n\right)^{1/3}
\end{align}

\section{Methodology}

We present a two-part computational framework designed to simulate dark matter–admixed neutron stars and to investigate their potential imprints on gravitational wave (GW) signals from binary neutron star (BNS) inspirals. Currently, the \lal{} framework, a comprehensive software suite developed by the LIGO–Virgo–KAGRA (LVK) Collaboration for gravitational wave simulations and data analysis~\cite{lalsuite}, does not provide support for neutron stars containing non-standard components such as DM. To address this limitation, we introduce \texttt{Darksuite}, a lightweight and extensible Python-based code that extends the \lal{} capabilities to include two-fluid stars with DM contributions. 

The codebase is implemented entirely in Python and relies only on essential libraries for numerical integration, interpolation, and plotting. Most functions in \texttt{Darksuite} are built from scratch to ensure maximum control and customizability. We aim to make \texttt{Darksuite} highly adaptable and easy to extend for future applications, including alternative DM models or modified gravity theories. 

\texttt{Darksuite} is organized into two main modules. The first module computes the equilibrium configuration of dark matter–admixed neutron stars based on chosen equations of state (EOS) and specified DM fractions. Key physical outputs include the total mass $M$, radius $R$, dimensionless tidal deformability $\Lambda$, and the mass fraction $f$ of DM. These quantities are evaluated over a grid of central densities to build a robust dataset. The second module performs interpolation over this dataset and uses the resulting stellar parameters to generate gravitational waveforms, enabling direct comparisons between signals from standard neutron stars and those with embedded DM.

The equilibrium configurations are obtained by numerically integrating the TOV equations. For ordinary neutron stars, the standard TOV system includes differential equations for the enclosed mass $m(r)$, pressure $p(r)$, and a perturbation function $y(r)$ used in computing the tidal deformability. For two-fluid stars, additional equations are included for the mass and pressure of the second component. We solve the coupled system using a fourth-order Runge–Kutta method \cite{RK4}. The equations are recast into a single vectorized ODE system for efficient and simultaneous integration.

As the physical radius $r_0 = 0$ leads to singularities in the differential equations, we initialize the integration from a small nonzero radius, e.g., at $r_0 \sim 10^{-6}$ km, which is negligible compared to a typical neutron star radius of $\sim 10$–$20$ km. The integration ends when the pressure of either fluid approaches zero and its mass stagnates. Multiple stopping criteria are implemented to improve reliability and capture edge cases, such as when one fluid halts earlier than the other (resulting in either a DM core or a DM halo). 

To support a wide range of models, we define a flexible EOS \texttt{class} that accepts an analytic parametric form of EOS as an argument. This class can generate plots and allow interpolation between pressure, energy density, and their derivatives on demand. Currently, the implementation supports three EOS models: nuclear matter, self-interacting bosonic DM, and fermionic DM. Future extensions can incorporate other exotic components by simply adding more objects of this EOS class.

A key challenge in simulating dark matter–admixed stars is determining the central energy density of the DM component that yields an expected $f$ with given NM central energy density. Since the final DM fraction depends on the central densities of both fluids in a non-linear way, it cannot be known a priori. We solve this inverse problem using a custom root-finding algorithm inspired by gradient descent. The algorithm starts with an initial guess for the DM's central density and iteratively adjusts it to minimize the difference between the resulting and desired $f$. Unlike conventional solvers, our algorithm is designed for interpretability and flexibility. It is straightforward, allowing tight control over the convergence process and making it easier to diagnose convergence issues. However, it also means that the solver is not fully optimized and may require further refinement to improve speed and stability.

In the second module, we interpolate $(C, f, k_2)$ data to compute $\Lambda$ for stars of arbitrary DM fractions. In particular, we leverage the \texttt{lalsimulation} module within \lal{}, which offers a broad set of waveform models for compact binary coalescence (CBC) signals. The use of \lal{} is motivated by its demonstrated success in analyzing previous gravitational wave catalogs~\cite{2019PhRvX...9c1040A, KAGRA:2021vkt, GWTC3, GW170817, GW190425, GW190814}, and its compatibility with other data analysis toolkits such as \bilby{}, \pycbc{}, and \gwpy{}.

To investigate the effects of dark matter on neutron star tidal interactions, we consider two representative DM equations of state (EOS): one for a degenerate fermionic species and one for a self-interacting bosonic condensate. For each EOS, we construct equilibrium stellar models by solving the two-fluid TOV equations over a two-dimensional grid in compactness $C = GM/(Rc^2)$ and dark matter mass fraction $f$. For each configuration, we compute the dimensionless tidal Love number $k_2$, which quantifies the star’s response to an external tidal field.

The resulting set of $(C, f, k_2)$ triplets defines a surface from which we construct a linear spline interpolation~\cite{jones2001scipy}. This enables efficient estimation $k_2$ for arbitrary stellar parameters within the grid, eliminating the need to reintegrate the structure equations for every new configuration. Given a compactness and DM fraction, we interpolate to determine $k_2$ and compute the corresponding tidal deformability $\Lambda$, which encodes the star’s quadrupolar response to tidal forces and impacts the inspiral phase phase of the gravitational waveform.

With $\Lambda$ determined, we compute the associated phase correction $\delta \Psi$ from \texttt{NRTidalv3}~\cite{Abac:2023ujg}, and use the \texttt{IMRPhenomPv2} waveform model~\cite{Khan:2018fmp} as implemented within \texttt{lasimulation} to generate the gravitational waveform strain $h(t)$, incorporating the tidal effects of the mixed dark matter–baryonic structure. We choose \texttt{IMRPhenomPv2} due to its extensive usage in \gw{} data analysis~\cite{2019PhRvX...9c1040A, Abbott2021Pop}. 
All simulations assume non-spinning, quasi-circular binaries to isolate the structural imprint of dark DM. This setup provides a computationally efficient approach to include dark-matter-induced tidal effects in gravitational waveform generation.

\texttt{Darksuite} provides a foundation for exploring the interplay between neutron star structure and gravitational waves in the presence of DM. While this initial version focuses on static equilibrium and simplified DM models, future versions will address current limitations, including computational efficiency, numerical robustness, and waveform post-processing.

\section{Results}
In this section, we describe the results of the TOV simulations of dark matter-admixed neutron stars with varying DM fractions, considering both bosonic and fermionic DM models. Afterwards, we analyze the gravitational wave signals induced from the simulated neutron star binaries, using parameters consistent with GW170817.

\begin{figure}\phantomsection
    \includegraphics[width=0.9\linewidth]{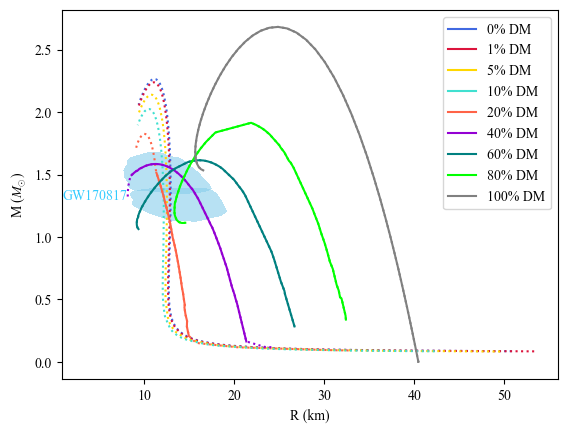}
    \includegraphics[width=0.9\linewidth]{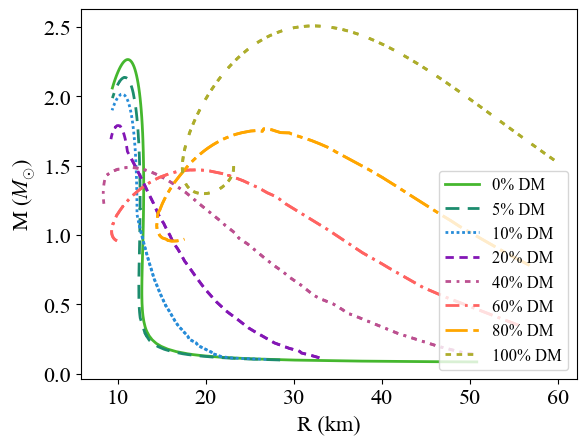}
    \caption{Mass \text{--} Radius relations for admixed neutron star of different dark matter ratios. The pure nuclear matter curve in both cases is modeled using the Brussels-Montreal functional BSk22. 
    \textbf{Top panel}: Self \text{--} interacting bosonic dark matter with $\rho_B \hbar = 9.1\times 10^{-4} \text{GeV}^4$. 
    \textbf{Bottom panel}: Fermionic dark matter with $\rho_F \hbar = 1.9\times 10^{-4} \text{GeV}^4$. Dashed lines represent stars with a dark matter core, whereas the solid lines correspond to a dark matter halo. The blue shaded region indicates the observational constraints from GW170817 (with 90\% CL).
    } 
    \label{fig:MR}
    
\end{figure}

FIG. \ref{fig:MR} presents the mass–radius stability curves for admixed neutron stars containing various DM fractions $f$, overlaid with observational constraints from the LIGO–Virgo detections of GW170817. These constraints are shown as credible intervals 90\% for the masses and radii of the components. All curves are generated using an identical set of central NM densities, except for the 100\% DM case, allowing for a consistent comparison between different DM admixtures. Dashed lines represent stars where DM is concentrated in the core, while solid lines indicate configurations with a surrounding DM halo. Nuclear matter is modeled using the Brussels–Montreal energy density functional BSk22. \textbf{In the top panel}, the DM component assumes a bosonic nature, characterized by $\rho_B\hbar^3 = 9.1 \times 10^{-4}~\text{GeV}^4$, deliveringng a maximum mass of approximately 2.7 $M_\odot$. \textbf{In the bottom panel}, a fermionic gas with the energy density scale of $\rho_F \hbar^3 = 1.9 \times 10^{-4}~\text{GeV}^4$ is used, corresponding to a particle mass of $0.49\,\text{GeV}$ and creating a maximum mass of 2.5 $M_\odot$. These parameters are chosen so that stellar objects of pure DM are of solar mass scale (following~\cite{Leung:2022wcf}).

Overall, in both panels, the presence of DM significantly alters the structural properties of the stars, resulting in a continuous evolution from NM-dominated to DM-dominated stability curves. At low DM fractions (5\% and 10\%), the mass-radius curves remain similar to that of the pure NM configuration, which is characterized by a hook-shaped segment at small radii, followed by a steep vertical decrease and an elongated horizontal tail near the end. However, significant structural changes begin to emerge from the 20\% and 40\% DM curves in both panels, where distinct cusps appear in the curves. The discontinuity corresponds to the transition of different internal configurations (i.e. either core or halo), which is visually indicated by changes in line style.

Specifically, the 20\% curve in the bosonic case exhibits two cusps, one at approximately 12 km and the other near 15 km. The segment to the left of the first cusp resembles the hooklike structure seen in the pure NM curve, as shown by the dashed line (representing the DM core). The portion between the two cusps mirrors the behavior of the pure DM curve, as interpreted by the solid line (DM halo). Beyond the second cusp, the curve again resembles the extended tail characteristic of the pure NM curve. Cusps clearly mark structural transitions that delineate whether the DM forms a central core or an extended halo. This distinction is reflected in the curve's morphology, that is, it behaves either NM-like or DM-like depending on the DM distribution. In reality, the transition points mark where the radii of the nuclear and DM components become equal. From the 60\% curve onward, the system transitions fully to a DM-dominated regime, at which point the hook and the tail features are no longer present. The same behavior is observed in the fermionic case.

The plotted curves are compared with posterior samples for the component stars of the BNS signal GW170817. The comparison reveals that the admixed DM models can reproduce configurations that lie within the observationally constrained regions, supporting the physical plausibility of such scenarios. However, we note that the posterior samples account for spin, whereas our TOV solver does not. Therefore, this comparison should be interpreted as a qualitative guideline rather than a precise quantitative match. A complete PE analysis using our algorithm for this event is beyond the scope of this work and is left for future study.

\begin{figure}\phantomsection
    \includegraphics[width=0.9\linewidth]{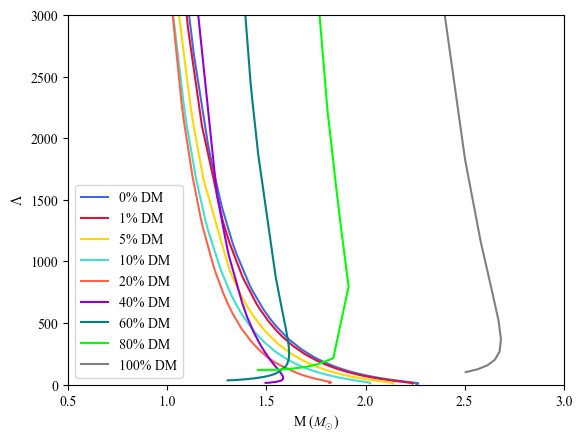} 
    \caption{Dimensionless tidal deformability-mass relations for admixed neutron star with varying $f$. The nuclear matter is modeled by the functional BSk22, while the dark matter component follows the self-interacting bosonic equation of state.}
    \label{fig: lambda-m}
\end{figure}

FIG. \ref{fig: lambda-m} shows the relationship between dimensionless tidal deformability $\Lambda$ and stellar mass $M$ for a range of DM fractions $f$. \textbf{In the top panel}, we used the self-interacting bosonic dark matter, whereas \textbf{In the bottom panel}, the DM component is fermionic. For low DM content (from 0\% to 20\%), the curves exhibit a consistent shape: a steep decline in $\Lambda$ at low masses, followed by a gentle slope towards the maximum mass. In addition, the curves progressively shift to the left of the baseline DM 0\% (pure NM) as $f$ increases within this range, indicating that stars with more DM become less susceptible to tidal deformation in the low DM region. 

On the other hand, a noticeable structural change between the 20\% and 40\% DM configuration marks a transition from the NM-dominated behavior to the DM-dominated behavior. Specifically, the 40\% DM curve differs significantly in shape, developing a pronounced hooked feature near the high-mass end, which is a characteristic shared by those beyond 40\% DM. Moreover, for the high DM content (from 40\% to 100\%) regime, the curves move in the opposite direction compared to the low DM regime, suggesting the type of stars that experience greater tidal deformation, rather than less.

Furthermore, the sensitivity of the curves to $f$ becomes more noticeable in regions with higher DM content. Although the difference between the curves 10\% and 20\% is subtle, the change from 80\% to 100\% is dramatic. This suggests that the system's response to additional DM is relatively modest at low DM fractions but becomes significantly more nonlinear in the high DM regime.

\begin{figure*}
    \centering
    \includegraphics[width=\linewidth]{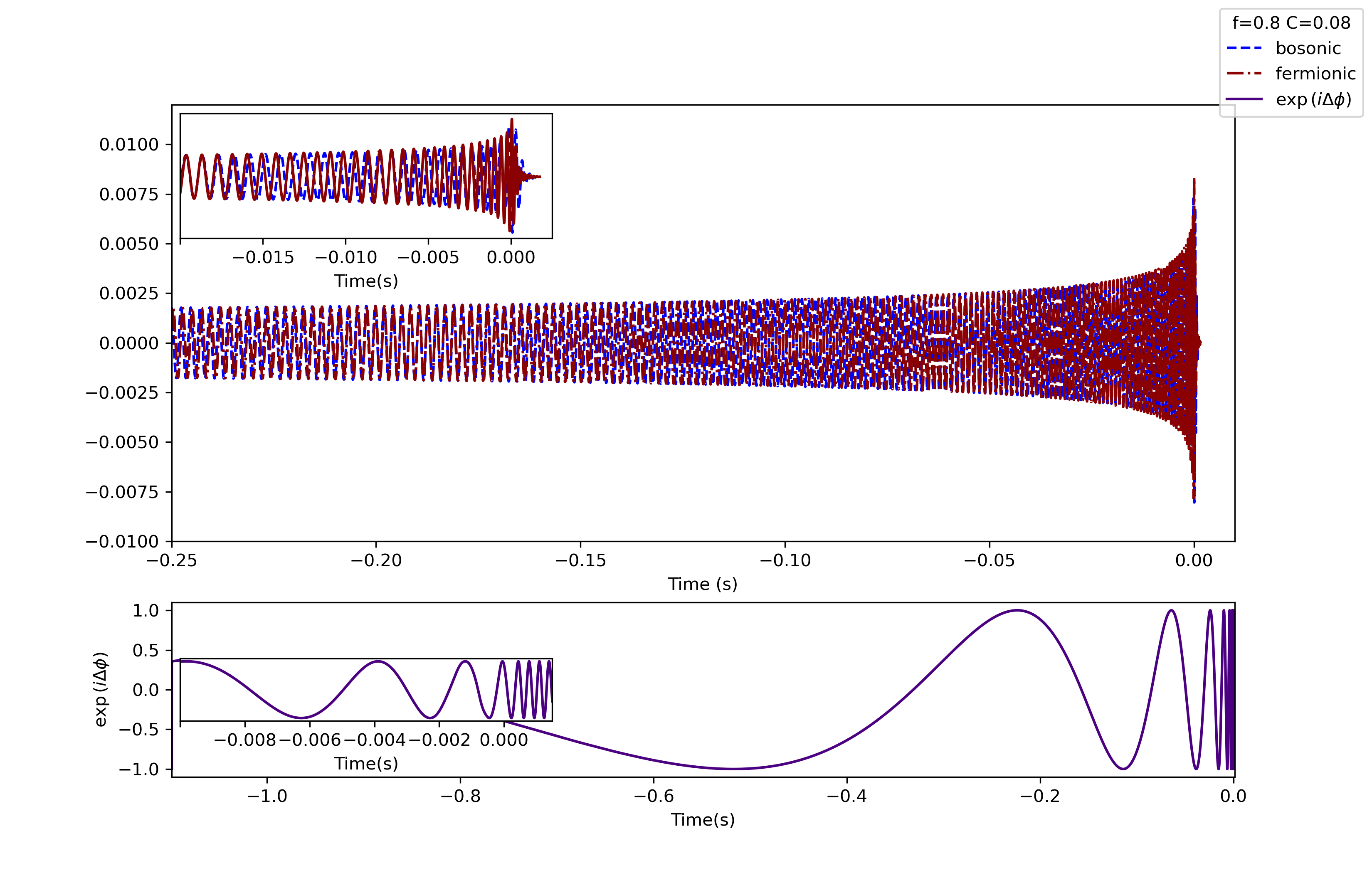}
    \caption{Gravitational waveforms in the case of bosonic (blue) and fermonic (dark red) dark matter, with $\rho_{B}\hbar^3 = 9.1 \times10^{-4} \ \mathrm{GeV^4}$ and $\rho_{F}\hbar^3 = 1.4 \times10^{-4} \ \mathrm{GeV^4}$ for the bosonic and fermionic cases, respectively. Waveforms are created with \texttt{NRTidalv3}, with the tidal deformability inferred form TOV simulations. The phase difference is plotted on the bottom pannel.}
    \label{fig:gws}
\end{figure*}

In Figure~\ref{fig:gws}, we plot two time-domain gravitational waveforms from binary neutron star (BNS) inspirals, each admixed with fermionic or bosonic dark matter. The simulation parameters are based on the BNS coalescence event GW170817, which consisted of two neutron stars with masses of 1.48 $M_\odot$ and 1.265 $M_\odot$. Both systems are constructed with identical compactness ($C = 0.08$) and a peak in tidal deformability at a fractional central dark matter density of $f = 0.8$, chosen to highlight the impact of dark matter composition on the gravitational waveform. The top panel shows the two waveforms for the full duration of the inspiral, generated using the \texttt{NRTidalv3} model within the \lal{}  waveform family. The inset in the top panel zooms in on the last $\sim$0.05 seconds of evolution, where the differences between the two waveforms become more pronounced.
The bottom panel shows the complex exponential of the phase difference between the bosonic and fermionic waveforms, $e^{i\Delta\Psi(t)}$, plotted as a function of time. This quantity emphasizes the accumulated phase shift over the inspiral and helps visualize when the waveforms go in and out of phase. The main panel covers the entire inspiral duration, while the inset zooms in on the final moments before the merger, where the phase discrepancy reaches its largest amplitude. The clear structure in $e^{i\Delta\Psi(t)}$ indicates that
phase evolution is modulated coherently throughout the inspiral, implying that even subtle dark matter effects can lead to measurable differences given a sufficiently high signal-to-noise ratio. This behavior is consistent with previous studies of dark-matter-admixed neutron stars, which have shown that even small differences in internal structure—such as those induced by varying dark-matter composition—can lead to cumulative phase shifts over the inspiral that become distinguishable near merger~\cite{Hinderer2009, Lackey2015, Ellis2018, Giangrandi:2025rko, Thakur:2024btu, Nguyen:2024fpq}

\section{Discussion}

The ability to resolve a phase difference between two gravitational waveforms that differ only by phase depends primarily on the signal-to-noise ratio (SNR), the detector's sensitivity across the relevant frequency band, and the duration of the signal within that band. At high SNR, typically above $\sim$ 20-30, even phase differences of a fraction of a radian can be resolved \cite{CutlerFlanagan1994}. This is especially relevant for long-lived signals such as those from binary neutron star (BNS) inspirals, where small phase shifts induced by physical effects like tidal deformability or dark matter interactions can accumulate over many cycles, leading to measurable deviations by the time the system merges \cite{FlanaganHinderer2008, Hinderer2009, Lackey2015}. Matched filtering techniques, commonly used in gravitational wave data analysis, are highly sensitive to phase differences, with the overlap between signal and template waveforms significantly decreasing as phase mismatches grow \cite{Owen1996}. Specifically, matched filtering allows discrimination of signals whose mismatch exceeds approximately \(1/(2\,\text{SNR}^2)\); at SNR~\(\approx 30\), this corresponds to a threshold of \(\sim 0.00056\) \cite{FlanaganHughes1998}. Even subtle differences in inspiral rate or spin-induced modulations become measurable under these conditions. The phase evolution of a gravitational waveform encodes critical source parameters such as chirp mass, mass ratio, and spins~\cite{Cutler1994, PoissonWill1995}. Bayesian parameter estimation further enables model comparison and precise recovery of source parameters, resolving even modest differences with high confidence at high SNR \cite{Veitch2015}. Although a two-detector network provides limited constraints on polarization and sky localization, it remains sufficient to detect variations in waveform phasing, particularly in the frequency domain where much of the signal power resides \cite{Abbott2016}. As illustrated in Fig.~\ref{fig:gws}, where the two waveforms go in and out of phase, such behavior implies a measurable difference in phase evolution, suggesting that, with current detector sensitivity and parameter estimation techniques, even waveform differences arising from exotic matter content such as bosonic or fermionic dark matter stars could be distinguishable.

The detectability of oscillatory phase modulations depends on the relationship between the modulation timescale and the signal frequency content. Taking GW170817 as a concrete example, the signal was observed with an SNR of approximately 30, its power concentrated near a central frequency of about 100 Hz, with the signal spanning a bandwidth of roughly 370 Hz over an observation duration of approximately 100 s. The phase difference between two waveforms can be expressed as a frequency-dependent complex factor $e^{i \Psi(f)}$, where $\Psi(f)$ is the phase modulation. Since $\Psi(f)$ is small, this factor can be expanded in a Taylor series as
$e^{i \Psi(f)} \approx 1 + i \Psi(f) - \frac{1}{2} \Psi(f)^2 + \ldots$. In first order, a sinusoidal form
$\Psi(f) = \Psi_0 \sin(2 \pi f T)$ is often used to model oscillatory phase modulations, where $\Psi_0$ is the modulation amplitude
and $T$ sets the characteristic frequency scale of oscillation. This form captures essential features of physical effects that induce periodic phase variations~\cite{FlanaganHinderer2008, Lackey2015}.  

Under this approximation, the mismatch between the two waveforms is approximately
$\mathcal{M} \approx \frac{1}{2} \Psi_0^2 \sin^2(2 \pi f_0 T)$, where $f_0 \approx 100\,\mathrm{Hz}$ is the central frequency.  
The detectability criterion, requiring $\mathcal{M} \geq 1/(2\, \mathrm{SNR}^2)$, gives a minimum detectable phase amplitude
$\Psi_0 \gtrsim \frac{1}{\mathrm{SNR} |\sin(2 \pi f_0 T)|}$. For GW170817 with $\mathrm{SNR} \approx 30$, phase modulations greater than
approximately 0.03 radians (about 2 degrees) are detectable when $\sin(2 \pi f_0 T)$ is near its maximum. When the sine term approaches zero,  
phase modulations become difficult to distinguish due to cancellation effects in the mismatch integral. If the modulation time scale $T$ is much shorter than
the inverse bandwidth $1/\Delta f \approx 1/370\,\mathrm{Hz} \approx 2.7\,\mathrm{ms}$, multiple oscillations occur in the frequency band
and the mismatch averages to $\mathcal{M} \approx \frac{1}{4} \Psi_0^2$, consistent with known scaling relations~\cite{CutlerFlanagan1994, Vallisneri2008}.  

If $\Psi(f)$ is not purely sinusoidal, it can be decomposed into sinusoidal components through Fourier analysis. Any realistic phase modulation
can be written as a sum of terms such as $\Psi(f) = \sum_n A_n \sin(\omega_n f + \phi_n)$, where each component has amplitude $A_n$ and frequency $\omega_n$.  
This approach allows one to quantify the sensitivity of current gravitational wave detectors to complex phase modulations without requiring full covariance matrix calculations. Although a complete calculation of this kind is beyond the scope of this paper, heuristically we can see that, again taking GW170817 as a concrete example, phase modulations varying much more slowly or much more rapidly than the inverse bandwidth of the signal produce phase differences that are nearly constant or average out across the frequency band, resulting in small mismatches; therefore, modulations with timescales comparable to the inverse bandwidth generate the largest mismatch and are most readily detected~\cite{CutlerFlanagan1994, Vallisneri2008, Owen1996, PoissonWill1995}.

\section{Conclusion}

In this article, we presented a program to systematically integrate dark matter into gravitational waveform modeling and assess its influence on neutron star mergers. This involves computing the tidal Love number for various equations of state (EoS), incorporating the dark matter ratio and particle type as key input parameters. Additionally, we determined the dimensionless tidal deformability parameters $\Lambda_1$ and $\Lambda_2$ using the calculated $k_2$ alongside the neutron star compactness $C = M/R$.
These parameters can be used to account for tidal corrections to gravitational waveforms that account for dark matter effects. We generated and analyzed waveform templates for dark matter-admixed neutron star binaries, comparing them to standard neutron star waveforms to identify distinct differences features. Finally, we extended to broader scenarios by incorporating alternative dark matter candidates such as self-interacting dark matter, as well as speculating on detectability and the effect of star rotation.

Ongoing estimation efforts \cite{Koehn:2024gal, KAGRA:2024ipf, Yuan:2021ebu}, particularly in the context of the next-generation Einstein telescope, have yielded mostly pessimistic results. However, the internal structure of dark-matter-admixed neutron stars can vary substantially depending on the microphysical properties of dark matter itself. In particular, the distinction between bosonic and fermionic dark matter—especially when self-interactions are included—can lead to measurable differences in the phase evolution of gravitational waveforms. These differences arise even when the overall mass and compactness of the star remain fixed, primarily because of variations in the density profile and tidal response. Our analysis shows that such phase modulations can accumulate coherently over the inspiral and may become detectable by current and future interferometers. These findings underscore the value of gravitational wave observations as a potential probe of dark matter microphysics, and motivate the continued development of waveform models that capture the full range of possible internal compositions.

\section{Acknowledgments}
This work is supported by a Phenikhaa University grant No. PU2024-3-A-05. Q.L.N. gratefully acknowledges support from the ICRR Inter-University Research Program.

\newpage

\bibliography{reference}
\end{document}